\documentclass[fleqn,usenatbib]{mnras}

\usepackage{newtxtext,newtxmath}

\usepackage[T1]{fontenc}
\usepackage{ulem}
\usepackage[dvipsnames]{xcolor}
\DeclareRobustCommand{\VAN}[3]{#2}
\let\VANthebibliography\thebibliography
\def\thebibliography{\DeclareRobustCommand{\VAN}[3]{##3}\VANthebibliography}
\newcommand{\kms}{\mbox{$\>{\rm km\, s^{-1}}$}}

\newcommand{\kpc}{\mbox{$\>{\rm kpc}$}} 
 
\newcommand{\Gyr}{\mbox{$\>{\rm Gyr}$}}

\newcommand{\kmsk}{\mbox{$\>{\rm kpc\, km\, s^{-1}}$}} 
\newcommand\degrees{^\circ}
\newcommand{\avg}[1]{\mbox{$\left<{#1}\right>$}}

\newcommand{\gaia}{{\it Gaia}}

\defcitealias{KH22}{K22}

\usepackage{graphicx}	
\usepackage{amsmath}	

\title[Pattern speeds of breathing waves]{The pattern speeds of vertical breathing waves}

\author[T. Khachaturyants et al.]{
Tigran Khachaturyants,$^{1}$\thanks{E-mail: astrotkh@gmail.com}
Victor P. Debattista,$^{1}$
Soumavo Ghosh,$^{2}$
Leandro {Beraldo e Silva},$^{3}$
\newauthor Kathryne J. Daniel$^{4,5}$
\\
$^1$Jeremiah Horrocks Institute, University of Central Lancashire, Preston PR1 2HE, UK\\
$^{2}$ Max-Planck-Institut f\"{u}r Astronomie, K\"{o}nigstuhl 17, D-69117 Heidelberg, Germany\\
$^3$Department of Astronomy, University of Michigan, 500 Church St., Ann Arbor,
MI, 48109, USA\\
$^4$ Department of Astronomy \& Steward Observatory, University of Arizona,
Tucson, AZ 85721, USA\\
$^5$ Department of Physics, Bryn Mawr College, Bryn Mawr, PA 19010, USA \\
}

\date{Accepted XXX. Received YYY; in original form ZZZ}

\pubyear{2022}

\begin{document}
\label{firstpage}
\pagerange{\pageref{firstpage}--\pageref{lastpage}}
\maketitle

\begin{abstract}
We measure and compare the pattern speeds of vertical breathing, vertical bending, and spiral density waves in two isolated N-body$+$SPH simulations, using windowed Fourier transforms over $1 \Gyr$ time intervals. We show that the pattern speeds of the breathing waves match those of the spirals but are different from those of the bending waves. We also observe matching pattern speeds between the bar and breathing waves. Our results not only strengthen the case that, throughout the disc, breathing motions are driven by spirals but indeed that the breathing motions are part and parcel of the spirals.
\end{abstract}

	\begin{keywords}
		stars: kinematics and dynamics --
		Galaxy: kinematics and dynamics --
		galaxies: disc --
		galaxies: evolution
	\end{keywords}



\section{Introduction}
It is now well known that the stars in the Solar neighbourhood and beyond exhibit prominent, large-scale coherent vertical motions, which can be decomposed into bending (coherently upwards or downwards) and breathing (coherently towards or away from the midplane) motions, as revealed by the second \gaia\ data release \citep[hereafter \gaia\ DR2,][]{Katzetal2018}, as well as by the LAMOST \citep{Carlin.etal.2013}, RAVE \citep{Williams.etal.2013}, and SEGUE \citep{Widrow.etal.2012} data. Since in an unperturbed, axisymmetric potential, the bulk vertical motion of stars is zero \citep{BT08}, the presence of such large-scale vertical motions points to some perturbing force driving these bulk motions. 
\par
Bending and breathing waves have been connected to several distinct dynamical processes. Bending motions can be generated via an interaction with a satellite \citep{HunterandToomre1969,Araki1985,Weinberg1991,sgr_ibata,sgr_dehnen,Gomezetal2013, Widrowetal2014, Donghiaetal2016,Schonrich+2018, binney+18,Chequersetal2018,laporte2019,Li&Shen2020,Helmi2020,Bennett+21,Poggioetal2021}, by the action of a buckling bar \citep{khoperskov2019}, or by irregular gas inflow along warps \citep[][hereafter \citetalias{KH22}]{KH22}. Vertical breathing motions are often attributed to the action of spiral density waves. On entering the spiral, stars are pulled by the density excess towards the midplane, and rebound upwards as they leave it \citep{Faureetal2014, Debattista2014, Monarietal2016, Ghoshetal2022,Kumar+22}. The amplitude of spiral-driven breathing motions is expected to increase with height from the midplane \citep{Debattista2014,Ghoshetal2022}, as is found for the breathing motions in the Milky Way \citep[MW,][]{Katzetal2018}. The relative sense of the breathing motion, whether towards or away from the midplane, is predicted to change across the spiral corotation resonance \citep{Faureetal2014, Debattista2014}. The change in direction emerges as compression (rarefaction) motions occurring more at the trailing (leading) edges of spirals within corotation and vice versa outside of corotation. \cite{Ghoshetal2022} found that, in their simulation, the spiral-driven vertical breathing motions decreased in amplitude with stellar age. They also observed this trend in the breathing amplitude using \gaia\ DR2 data, indicating that spiral density waves might well drive the breathing motion in the MW.
Analytical studies have shown that bars can also drive breathing motions \citep{Monarietal2015}, while \citet{Widrowetal2014} have suggested that tidal interactions may also do so.

In this Letter, we explore the nature of breathing waves by measuring, for the first time, their pattern speeds in two isolated high-resolution $N$-body+SPH simulations. We also measure the pattern speeds of vertical bending motions and of spiral density waves, and compare their evolution with those of the breathing motions. We demonstrate that the spirals and the breathing waves share the same pattern speeds: when a density pattern speed with significant power is present in the disc, we find a matching breathing wave and, conversely, when a breathing pattern speed with significant power is present, we find a matching density perturbation (spirals or a bar). We conclude that breathing waves are merely the vertical extension of spirals or bars, with no separate lives of their own.


\section{Simulations}

We use the two $N$-body+SPH simulations described in \citetalias{KH22} which are evolved using the code \textsc{gasoline} \citep{Wadsley+2004} for $12 \Gyr$ without any mergers. In one model, the dark matter halo is spherical and, therefore, the accretion of cooling gas from the corona is mostly in-plane with the disc, resulting in a relatively flat disc. In the second model, the dark matter halo is triaxal with the angular momentum of the embedded gas corona misaligned with its principal axes; this results in gas accreting via an S-shaped warp onto the disc. Hereafter, we refer to these two models as the unwarped and warped models. \citetalias{KH22} demonstrated that misaligned gas accretion generates bending waves with amplitudes larger than those in the unwarped model. Detailed discussion of the initial conditions and evolution of the two models is presented in \citetalias{KH22}.


\section{Spectral analysis}

In \citetalias{KH22}, we showed that both simulations support density waves (spirals) and bending waves throughout their evolution. Bending waves are stronger in the warped model but are also present in the unwarped model. 
We use standard measures for bending ($v_{+}$) and breathing ($v_{-}$) velocities which are defined as:
\begin{equation}
v_\pm(x,y,z) = \frac{1}{2}\left[ \avg{v_z(x,y,+z)} \pm\avg{v_z(x,y,-z)} \right].\\
\label{eq:bnd_eq}
\end{equation}
The $v_z$ averages are computed over stars above and below the midplane in a given vertical column. The simultaneous presence of bending and breathing waves requires that, in measuring the breathing velocity, we compute the average velocity field to decouple the orthogonal motions. We achieve this by computing the binned \avg{v_z} on a regular grid. With this method we are able to measure the vertical motions' pattern speeds independent of each other and of the density waves. Instead, a particle-based measurement, as usually done for spiral waves, is not possible for breathing waves.

\subsection{Binning technique}
\label{sec:bin}
We bin our data in cylindrical coordinates with $R\in[0,10] \kpc$ and $\Delta R = 0.5 \kpc$, $|z|\in[0.3,1.5] \kpc$ and $\Delta z = 0.3 \kpc$, and $\phi\in[0,360] \degrees$ and $\Delta \phi = 15\degrees$. We cut out the midplane as breathing signals are expected to increase with height from the midplane; we confirm that this does not affect the measurement of the bending and density waves. In the binned space, we compute the stellar mass, $m_*(R_i,\phi_j,z_k)$ and mean vertical velocity, $\avg{v_z(R_i,\phi_j,z_k)}$.
The binning produces a 3D cylindrical array which is then processed based on the wave being analysed. For density waves, the mass array is collapsed onto the midplane by summing all bins over the $z$-axis:
\begin{equation}
	M(R_i,\phi_j)= \sum_{k=0}^N m_*(R_i,\phi_j,z_k)
\end{equation}
thereby producing a $2$D polar array. For bending and breathing waves, the procedure requires a separation between $\avg{v_z}$ values above and below the midplane; we separately collapse bins above and below the midplane by summing $\avg{v_z(z>0)}$ and $\avg{v_z(z<0)}$ along the $z$-axis, respectively. The resulting arrays are then subtracted or summed to measure the breathing ($V_{-}$) or bending ($V_{+}$) motions, respectively:
\begin{equation}	
		V_{\pm}(R_i,\phi_j)=
	\frac{1}{2}\sum_{k=0}^{N/2}\left[ \avg{v_z(R_i,\phi_j,z_k)}\pm\avg{v_z(R_i,\phi_j,-z_k)}\right],
	\label{eq:br_2d}
\end{equation}

\subsection{Pattern speed calculation}
We modify the spectral analysis code of \citetalias{KH22} to work with binned data. We use these binned data to measure the pattern speeds of the density, bending, and breathing waves. After producing the $2$D polar arrays described in Section~\ref{sec:bin}, we calculate the Fourier coefficients by assuming that each element of the array represents a particle with mass $M(R_i,\phi_j)$ and vertical velocity $\avg{v_z(R_i,\phi_j)}$. The Fourier coefficients for density waves are computed as
\begin{equation}
	c_m(R_i)=\frac{1}{\sum_{j}^{}M(R_i,\phi_j)} \sum_{j} \left[ M(R_i,\phi_j) ~ \rm{e}^{im\phi_j} \right],
\label{eq:dens}
\end{equation}
where $c_m(R_i)$ is proportional to the surface density contrast, $\Delta\Sigma/\Sigma$. Coefficients for breathing ($V_{-}$) and bending ($V_{+}$) waves are derived in a similar way, but to avoid any direct influence of the density waves on the breathing/bending signal, the mass distribution is factored out and we instead normalise by the number of azimuthal bins, $N_\phi$ (the results are not affected by the inclusion of mass, as in Eq. \ref{eq:dens}):
\begin{equation}
	\gamma_{m,\pm}(R_i)= {N_\phi}^{-1}\sum_{j} ~V_{\pm}(R_i,\phi_j) ~ \rm{e}^{im\phi_j}.
\label{eq:br}
\end{equation}
We focus on the positive pattern speeds, $\Omega_{\rm{p}}$, of density, bending, and breathing waves, so the method described in \cite{Roskar+12} and the coefficients calculated in Eqs.~\ref{eq:dens} and \ref{eq:br} are used to obtain $\Omega_{\rm{p}}$. 

While the particle-based method of \citetalias{KH22} uses the $z$ coordinate to obtain the Fourier coefficients for bending waves, here we use $V_+$ which is based on $v_z$. This change is necessary since the current analysis requires binning the data, and computing $\avg{z}$ in a given bin would effectively output the centre of each bin. We first confirm that using (unbinned) $v_z$ instead of $z$ does not affect the pattern speed of the bending waves. We repeat the unbinned analysis of \citetalias{KH22} for both $z$ and $v_z$ and confirm that the total power and pattern speeds of the bending waves based on $v_z$ match those based on $z$.
 
Additionally, to verify that the binning does not bias our results, we compare the pattern speeds computed for the binned density and bending waves with those of the particle-based method of \citetalias{KH22}. The comparison of the most prominent pattern speeds in each baseline yields Pearson correlation coefficients of $0.93-0.95$ for both density and bending waves, demonstrating a good agreement between the binned and particle methods. Only in the weaker spirals and bending waves is the difference between the binned and particle-based pattern speeds larger. We conclude that the binning method is in good agreement with the results of \citetalias{KH22} and, therefore, suitable for our analysis of the pattern speeds of density, bending, and breathing waves.

\section{Results}
\label{sec:results}
\begin{figure*}
    \centering
    \includegraphics[width=1\linewidth]{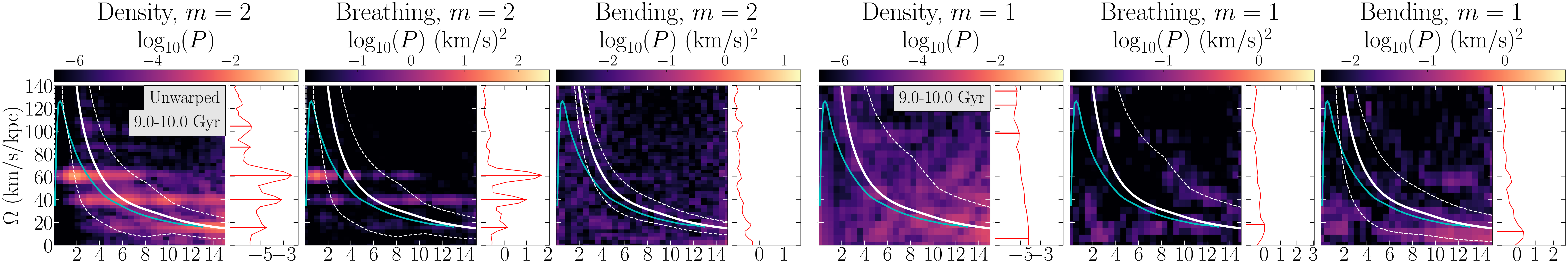}
    \centering
    \includegraphics[width=1\linewidth]{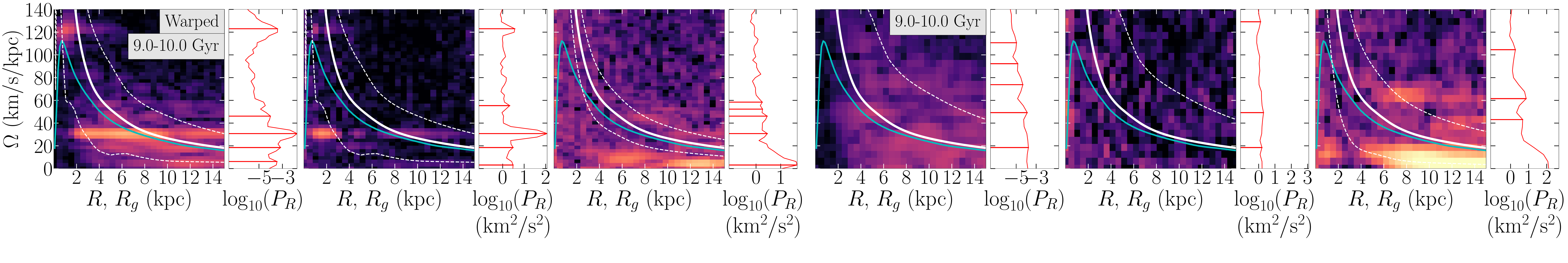}
    \caption{Power spectra (spectrograms) for the $m=2$ (left block) and the $m=1$ (right block) waves in the unwarped (top) and warped (bottom) simulations at a time interval which contains the strongest density wave. The white solid lines show the rotation curve, $\Omega(R)$, evaluated at the central time of the interval using an interpolated potential \citepalias{KH22}. White dashed lines in the density and breathing columns indicate the inner/outer Lindblad resonances, $\Omega\pm\kappa/m$, while the dashed lines in the bending wave column indicate the vertical resonances, $\Omega\pm\nu/m$. The cyan lines show the rotation curves, $\Omega_\varphi(R_g)$, for vertically hot stars ($z_{max}\geq1.5\kpc$). The first column of each row shows the power spectra for density (bar$+$spiral) perturbations. The second and third columns of each row show the power spectra for breathing and bending waves, respectively. To the right of each spectrogram we show the radius-integrated power. A peak-finding algorithm is applied to each total-power curve with the detected peaks indicated by the horizontal red lines. The colour scale limits are the same for each respective wave type over all multiplicities and models. The pattern speeds of ($m=2$) density and breathing waves match each other, but not the bending waves.
}
    \label{fig:739HF_hc_psp}
\end{figure*}
Fig.~\ref{fig:739HF_hc_psp} (left blocks) shows the spectrograms and power spectra for the $m=2$ density, breathing, and bending waves (indicated at the top of each column) in the unwarped and warped simulations (see annotations) at a time interval containing the strongest density wave. Although, for brevity, we only show one time interval, we have examined the entire evolution of both models. The solid white lines show the rotation curves, $\Omega(R)$. The dashed white lines in the density and breathing columns indicate the inner/outer Lindblad resonances, $\Omega \pm \kappa/m$, and the vertical resonances, $\Omega \pm \nu/m$ in the bending columns. By radially integrating the spectrograms, we produce the total power spectra, $P_R(\Omega)$, shown to the right of each spectrogram panel (red curves). In both models, we note a striking resemblance in the shapes and peaks of the density and breathing power spectra, while the spectrograms for the bending waves are very different. We also note the presence of a small bar in the unwarped ($\Omega_{\rm{p}}\sim60\kmsk$) and warped ($\Omega_{\rm{p}}\sim120\kmsk$) simulations, with a pattern speed signal extending beyond co-rotation, suggesting a resonantly coupled spiral. There is a match between the bar and one of the breathing waves in both models, which is in agreement with \cite{Monarietal2015}. We apply a peak finding algorithm \citep{scipy} and indicate the most prominent and strongest peaks with horizontal red lines. Based on the total power distributions, there is an even clearer correlation between the peaks of the density (bar$+$spiral) and the breathing waves, while the bending waves show no correlation with either the density or breathing waves.

In \citetalias{KH22}, we showed that most of the bending wave power is concentrated at $m=1$. Therefore, in Fig.~\ref{fig:739HF_hc_psp} (right blocks) we present the spectrograms and power spectra for the $m=1$ density, breathing, and bending waves in the two simulations. The $m=1$ spirals and breathing waves are significantly weaker than in $m=2$. As a consequence, the matches between density and breathing waves become less apparent, with only some intervals having peaks with matching pattern speeds. Unlike the visibly pervasive matches between the $m=2$ density and breathing waves, we only observe a few sporadic matches between breathing and bending waves. Additionally, the stronger bending waves in the warped model nonetheless remain poorly matched to the breathing waves.

The spectrograms of $m=2$ breathing waves in both simulations exhibit a gap for any given pattern speed near its corotation. This agrees with the need for stars to transit the spiral and be pulled by the spiral's excess density. While the rotation curves in Fig.~\ref{fig:739HF_hc_psp} are for planar circular (cold) orbits, the angular frequency of hotter populations should be smaller at a given radius. Using \textsc{agama} \citep{agama}, we compute the guiding radius, $R_g$, and the angular frequency, $\Omega_{\varphi}$, of vertically hot stars ($z_{\rm{max}}\geq1.5\kpc$), which we show as cyan lines (averaging over stars at fixed $R_g$). The resulting (hot) corotation radii match the gaps better, since stars further from the midplane contribute  the most to $V_-$ \citep{Faureetal2014, Debattista2014}.

\begin{figure}
    \centering
    \includegraphics[width=1.\linewidth]{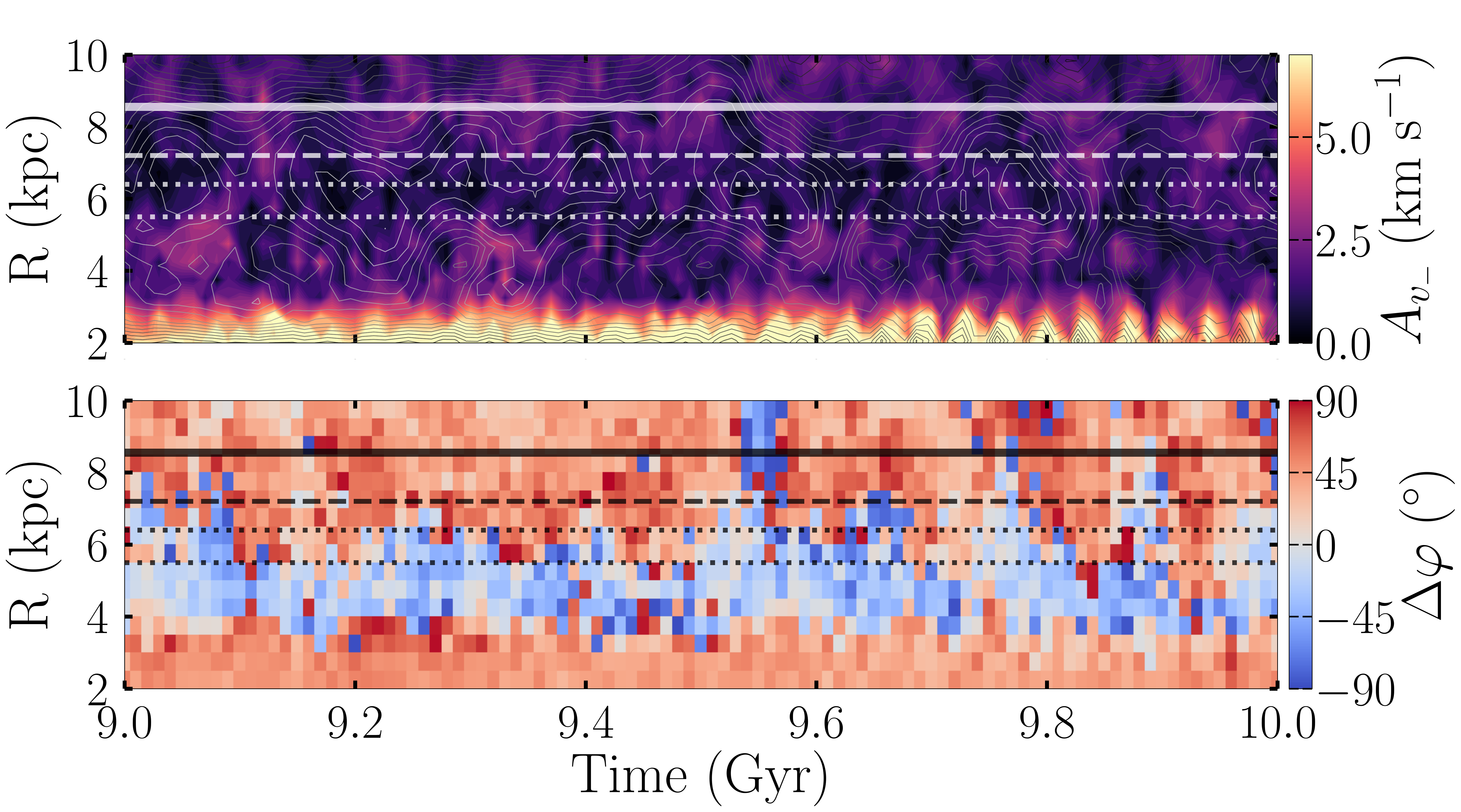}
    \caption{Evolution of the breathing (top, colour) and density (top, contours) wave amplitudes and of the relative phase, $\Delta \varphi$, between the two waves (bottom, colour) in the warped model at the time interval in Fig.~\ref{fig:739HF_hc_psp}. The horizontal lines indicate the gap in Fig.~\ref{fig:739HF_hc_psp} where the $m=2$ breathing wave power drops (dotted) and the locations of the cold (solid) and hot (dashed) corotation radii. We see a clear separation between the bar, and across the (hot) corotation radius, of patches with approximately constant $\Delta\varphi$ over time, in agreement with breathing motions being spiral-driven.
    }
    \label{fig:phase}
\end{figure}

Fig.~\ref{fig:phase} presents the evolution of the breathing amplitude, $A_{v_-}$ (top), and the relative phase between the breathing and density perturbations, $\Delta \varphi$, (bottom), both computed using the Fourier coefficients of Eqs.~\ref{eq:dens} and~\ref{eq:br}. Multiple strong density waves are present in the unwarped model, which produce complex interference patterns, thus for illustrative purposes we show the warped simulation at a time where only one spiral is dominant. The lines indicate the cold (solid) and hot (dashed) corotation radii, while the dotted lines show the gap locations. Three distinct regions of strong breathing amplitudes are evident: within the bar, and in the inner and outer disc. The latter are separated by the hot corotation radius and the gap. The evolution of $\Delta \varphi$ exhibits three regions of near-constant values, mirroring those in $A_{v_-}$. Negative and positive $\Delta \varphi$ occur within and beyond the corotation radius, respectively, in agreement with spiral-driven breathing in the disc region.

\begin{figure*}
    \centering
    \includegraphics[width=1.\linewidth]{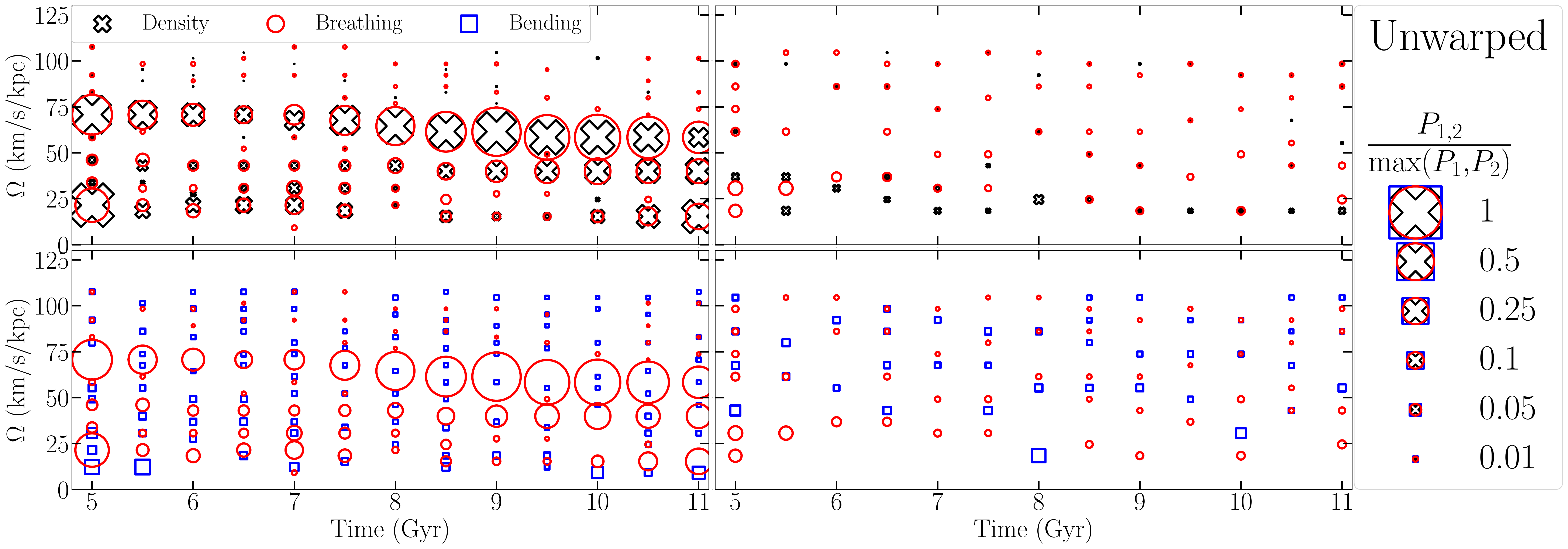}
    \centering
    \includegraphics[width=1.\linewidth]{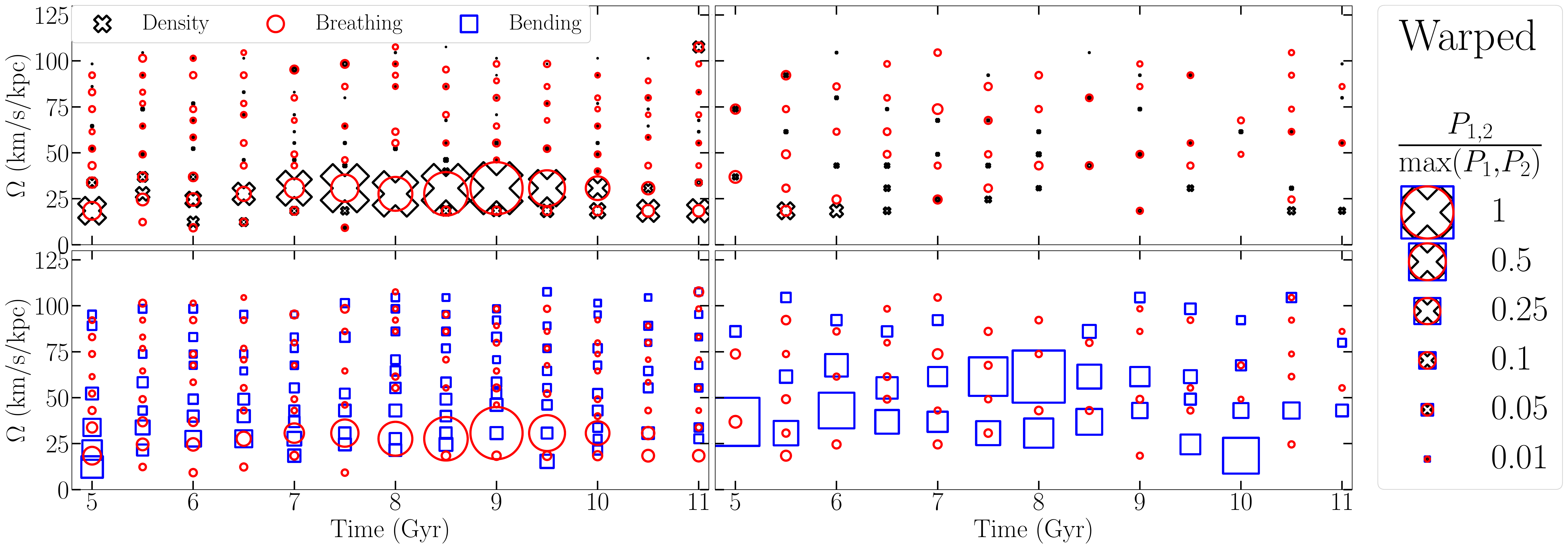}
    \caption{The most prominent pattern speeds in the unwarped (top) and warped (bottom) models identified in Fig. \ref{fig:739HF_hc_psp}, for the $m=2$ (left) and the $m=1$ (right) density (black crosses), breathing (red circles), and bending (blue squares) signals. The size indicates the peak height in the radius-integrated power at a given pattern speed. The peak height is scaled by the highest peak in the density ($10^{5.12}\,$), breathing ($10^{3.92}$ $\rm{km}^2\rm{s}^{-2}$), and bending ($10^{3.58}$ $\rm{km}^2\rm{s}^{-2}$) waves over both multiplicities and models, $P_m/\rm{max}(P_1,P_2)$. 
    The rightmost legend shows the relationship between the marker size and $P_{m}/\rm{max}(P_{1,2})$.
    We observe a correlation between the density and breathing signals, which is particularly strong for $m=2$; the breathing and bending signals show no correlation.}
    \label{fig:bubble_uw}
\end{figure*}
To illustrate the correlation between the different waves, Fig.~\ref{fig:bubble_uw} shows `bubble plots' \citep{Roskar+12} for the unwarped (top) and warped (bottom) simulations for $m=2$ (left) and $m=1$ (right). The pattern speeds are displayed as symbols with varying size, based on the strength of the identified peaks -- see  Fig.~\ref{fig:bubble_uw} for details. The evolution of the peak $m=2$ density (black crosses) and breathing (red circles) pattern speeds (top left panel) are, generally, in very good agreement at different times. While we find multiple matches between the density and breathing pattern speeds in $m=1$ (top right panel), these matches are less prevalent than in $m=2$.
The evolution of peak breathing and bending (blue squares) pattern speeds (bottom row) present no clear matches in either $m=1$ or $m=2$, although there are some time intervals where the pattern speeds overlap. Unlike the density and breathing perturbations, these pattern speed overlaps are sporadic and do not exhibit a clear trend.

\begin{figure}
    \centering
    \includegraphics[width=1.\linewidth]{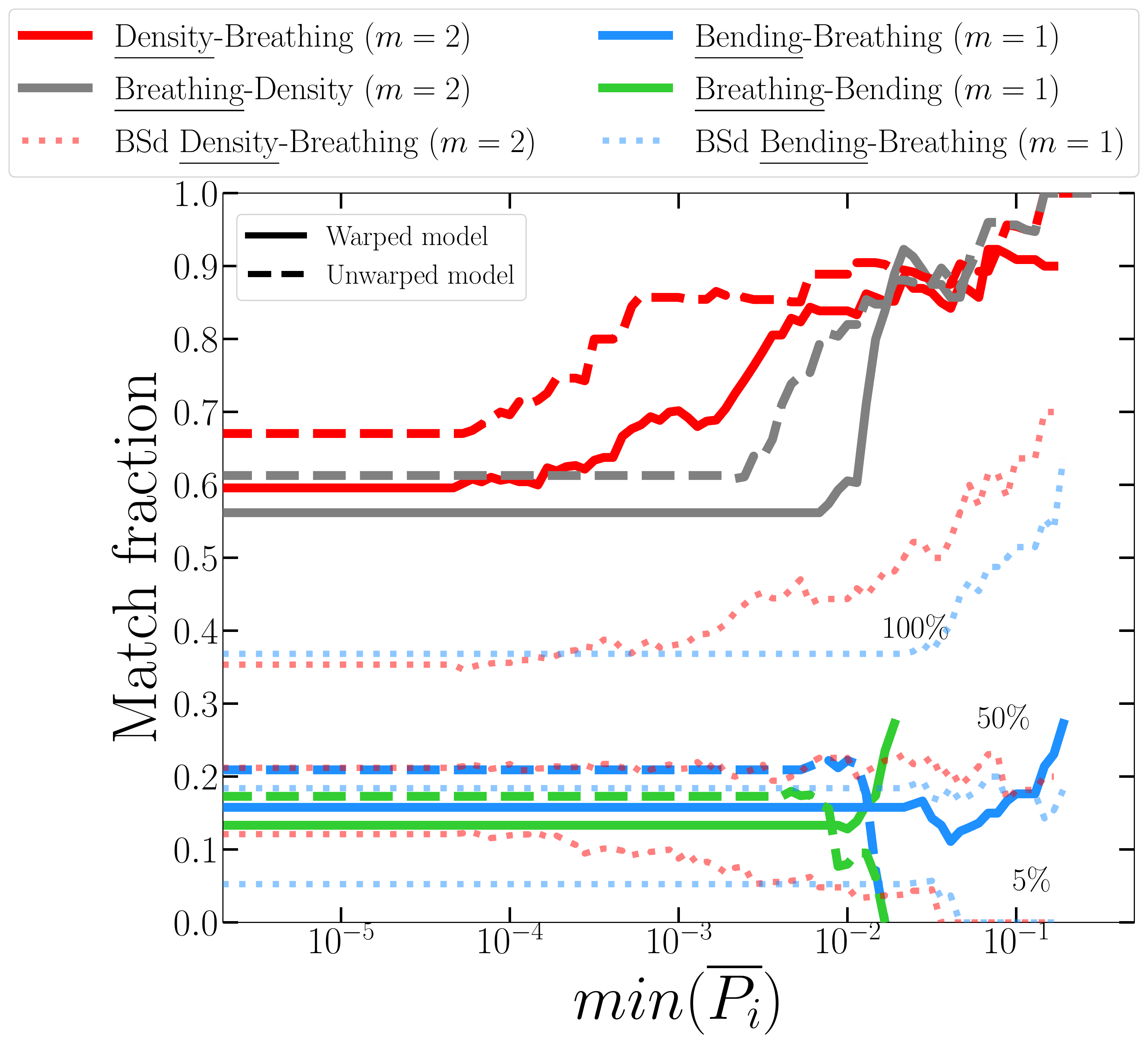}
    \caption{Match fraction between two waves, indicated in the legend, as a function of $\min(\overline{P_i})$, the lower cutoff of the normalised total power of the primary waves. 
    The matches are computed in the warped (solid) and unwarped (dashed) models over all times.
    The normalisation is based on the maximum values in Fig~\ref{fig:bubble_uw}. 
    The dotted lines indicate different percentiles of match fractions from 2500 random bootstrap samplings (BSd) over all baselines in the warped model (the results for the unwarped model are qualitatively very similar).  Increasing $\min(\overline{P_i})$ for $m=2$ density (red) and breathing (grey) waves results in an increase in the match fraction larger than the BSd random match fractions (the largest $\min(\overline{P_i})$ values contain at least 10 primary frequencies). The $m=1$ bending (blue) and breathing (green) waves show no significant change in match fraction with increasing $\min(\overline{P_i})$, and instead are consistent with the random matches of the BSd fractions.
    }
    \label{fig:match_freq}
\end{figure}

To quantify the significance of the match between the breathing and density/bending waves, in Fig.~\ref{fig:match_freq}, we show the fraction of identified pattern speeds in one wave type (primary wave) that are also identified in another wave type (target wave) at the same value. Specifically, we define the match fraction as the ratio of the number of matched primary pattern speeds to the total number of primary pattern speeds. We show this fraction for different minimum power thresholds of the primary pattern speeds, $\min(\overline{P_i})$. Thus at the lowest $\min(\overline{P_i})$, the match fraction indicates the matches for all primary pattern speeds. The total power is normalised by the strongest pattern speeds of that type identified in Fig.~\ref{fig:bubble_uw}. Stronger $m=2$ density and $m=2$ breathing waves (larger $\min(\overline{P_i})$), produce significantly larger match fractions: at $\min(\overline{P_i})=10^{-2}$ selecting $m=2$ density ($m=2$ breathing) as the primary wave, the match fraction with breathing (density) waves reaches $\sim90-95$ per cent ($\sim85-95$ per cent). 

On the contrary, increasing $\min(\overline{P_i})$ in the $m=1$ bending and breathing waves leads to little change in their match fractions of $\sim0.25$; however this relatively high fraction may suggest some connection between these two wave types. To check if these match fractions are compatible with random matches, we select 2500 random bootstrap samples of $m=1$ and $m=2$ breathing pattern speeds across all baselines in the warped model (the results for the unwarped model are qualitatively similar). We use the same spectral resolution ($\sim 6 \kmsk$ in $m=1$ and $\sim 3 \kmsk$ in $m=2$) as in our spectral analysis. The bootstrap-sampled pattern speeds are then matched with density and bending waves of corresponding multiplicities and the random match fractions $5$, $50$, and $100$ percentiles are displayed as dotted lines. Match fractions for $m=1$ bending and breathing waves (blue) reach the observed values often in these random resamples, with a $50\%$ probability of reaching $~\sim0.20$. Unlike the $m=1$ bending waves, the $m=2$ density-breathing bootstrapped sample (red) does not reach the observed match fractions at higher $\min(\overline{P_i})$, indicating that these matches are very significantly above the random level. Lastly, we test if the observed match fractions could be artificially inflated by a large number of target frequencies. By limiting the number of target wave frequencies to no more than in the primary waves, the match fraction between density and breathing waves remains the same, while the one between bending and breathing waves reduces further to $\sim0.15$.

\section{Conclusions}

We have compared the pattern speeds of breathing waves with those of spiral density waves and of bending waves. We have found that the pattern speeds of the breathing waves match those of the $m=2$ spirals (and a bar if one is present), provided the spirals are relatively strong. In contrast, bending waves and breathing waves propagate separately. This is true even when the bending waves are strong, as in the warped simulation. We also observe that breathing waves weaken near the corotation radii as would be expected if they are driven by density waves in the disc. We conclude that spirals are the main drivers of breathing waves. We do not find any strong spirals without accompanying breathing waves. Likewise, in almost all cases the breathing waves are accompanied by spirals at the same frequency. The breathing waves therefore do not have an independent existence but rather are part of the very structure of spirals.

\bigskip
\noindent
{\bf Acknowledgements.}

\noindent
V.P.D. and T.K. were supported by STFC Consolidated grant ST/R000786/1.  L.B.S acknowledges the support of NASA-ATP award 80NSSC20K0509 and Science Foundation AAG grant AST-2009122. The simulations in this paper were run on the DiRAC Shared Memory Processing system at the University of Cambridge, operated by the COSMOS Project at the Department of Applied Mathematics and Theoretical Physics on behalf of the STFC DiRAC HPC Facility (www.dirac.ac.uk). Analysis was carried out on Stardynamics, which was funded by the Newton Advanced Fellowship NA150272 awarded by the Royal Society and the Newton Fund.

\bigskip
\noindent
\section*{Data Availability}

\noindent
The simulation dataset used here can be shared for limited use on request to V.P.D. (vpdebattista@gmail.com).



\bibliographystyle{mnras}
\bibliography{example} 





\bsp	
\label{lastpage}
\end{document}